# Synthesis and characterization of few-layer GeSe by liquid phase exfoliation

Yuting Ye[a], Qiangbin Guo[a], Xiaofeng Liu*[a], Chang Liu[a], Junjie Wang*[c], Yi Liu[a], Jianrong Qiu*[b]

[a]School of Materials Science and Engineering, Zhejiang University Hangzhou, 310027, China
E-mail: xfliu@zju.edu.cn

[b]State Key Laboratory of Modern Optical Instrumentation, College of Optical Science and Engineering, Zhejiang University,          Hangzhou, 310027, China, E-mail: qjr@zju.edu.cn

[c]Materials Research Center for Element Strategy, Tokyo Institute of Technology
4259 Nagatsuta, Midori-ku, Yokohama, 226-8503, Japan
E-mail: wang.junjun0810@gmail.com

**Abstract:** Monochalcogenides of germanium (or tin) are considered as stable isoelectronic and isostructural analogue of black phosphorous. Their two-dimensional (2D) forms have been just predicted to shown strong thickness-dependent physical properties, and even indirect to direct band gap crossover at the monolayer limit. Here, we demonstrate the synthesis of atomically thin GeSe by direct sonication-assisted exfoliation of bulk microcrystalline powders in solvents. The exfoliated few-layer GeSe sheets characterize high crystallinity with lateral sizes over 100 nm and, importantly, strong resistance against oxidation and degradation in ambient conditions. Density functional theory calculation combined with optical characterizations confirm the layer-number dependent optical bandgap, which, for few-layer sheets, is optimal for solar light harvesting, and promising for relevant applications, such as optoelectronics and photonics.

Recent decade has witnessed the booming of extensive investigations into diverse two-dimensional (2D) materials of atomic-scale thickness since the discovery of graphene in 2004.[1] The near-zero electron effective mass associated with the liner energy dispersion endow graphene with extremely high electron mobility, promising for novel electronic devices. Its application is only limited by the gapless nature which has plagued the development of switching electronic devices. In comparison, 2D semiconductors based on transition metal dichalcogenides (TMDs) with bandgaps of 1.5 - 2.5 eV are considered better semiconducting 2D scaffold for future electronics, while they usually characterize smaller mobility associated with the relatively large bandgap and the small dispersion near the band edges.[2] The 2D form of black phosphorus (BP), as known as phosphorene, with a strong layer-dependent bandgap of 0.3 - 1.5 eV therefore becomes a better alternative 2D semiconductor, demonstrating with excellent electronic properties yet compromised stability compared with 2D TMDs.[3]

In the meantime, it is quite natural to explore BP's isoelectronic analogs, i.e., compounds with 10 electron (for each atom pair) for possible 2D semiconductors.[4] Indeed, the compounds of group IV-VI elements, such as the group IV monochalcogenides (i.e., GeS and GeSe), which have bandgaps of 1.0 - 2.3 eV, usually adopt a similar puckered layer structure as that of BP, while the difference is that the cations in GeSe (or GeS) project out of the layer (See Figure 1a,b). In addition, each cation in this structure are three-coordinated, leaving a long electron pair pointing also to the interlayer spacing, which, therefore, is sensitive to interlayer coupling. The bulk form of these compounds, including both chalcogenides and oxides, have in fact unraveled interesting properties, such as carrier bi-polarity,[5] superconductivity,[6] as well as high piezoelectric and thermoelectric performance.[7,8] Recent theoretical calculation for the mono- and few-layer (FL) form of these chalcogenides have shown the strong layer-dependent optical and electric properties as well as the valley polarized excitons due to breaking of inversion symmetry in monolayer.[9,10] Especially, GeSe, with a bulk bandgap close to that of silicon (1.12 eV), have been predicted to experience the indirect to direct bandgap crossover for monolayer,[4b,9b] similar as that of $MoS_2$.[11] On the other hand, the heavier atomic mass (as compared to BP) of these 2D compounds also makes them a better platform for exploiting spin-orbital physics.[12] These peculiar features have recently evoked growing concerns on these 2D IV-VI compounds like GeSe,[4-10] which, however, has not been accessed experimentally by existing process variable to other 2D materials.

We demonstrate here successful liquid phase exfoliation (LPE) of the GeSe with atomic thickness and high crystallinity. These FL-GeSe sheets are re-dispersible in various solvent forming stable colloidal solution with long term stability, and importantly, high robustness against degradation. First principle calculations and optical measurement both confirm clear layer-number dependent optical properties and we show that FL-GeSe could be a highly active 2D medium with remarkable photocurrent generation under visible light irradiation, which might be promising for energy-related and photonic applications.

GeSe adopts the orthorhombic structure and it characterizes a monolayer thickness of 0.54 nm with a van der Waals gap of 0.26 nm. To understand the role played by inter-layer coupling in optical and relevant properties, we first calculated the electronic structure for mono-, bi-layer and bulk GeSe (Figure 1d,e). Overall, the structural two-dimensionality is responsible for the very small dispersion along F-Q-B, as well as the presence of sharp peaks in density of states (DOS) which can be associated with strong optical absorption favorable for photovoltaic applications.[13,12] Similar as other 2D semiconductors (e.g., TMDs),[11] the bandgap expands drastically, from 0.87 eV for the bulk to 1.17 eV at the monolayer limit, in agreement with a recent report.[10] Interestingly, an indirect to direct bandgap crossover is also observed when GeSe is thinned down to monolayer, suggesting the strong influence of van der Waals interaction on electronic structures. In comparison, the minimum of conduction band (CB) and maximum of valence band (VB) are located at different *k*-points for both bulk and bi-layer GeSe. From the projected DOS (PDOS), the top of the VB and the bottom of the CB also undergoes significant modifications in the PDOS of both Ge and Se, as GeSe is covalent and therefore Ge and Se orbitals are highly hybridized. A more detailed analysis reveals that the bottom of the conduction band is mainly contributed from orbitals of both the Se 4p and Ge 4p, and importantly the Ge 4s (due to 4s long pair electrons in bivalent Ge), which may be responsible for the high sensitivity of electronic structure to layer stacking.

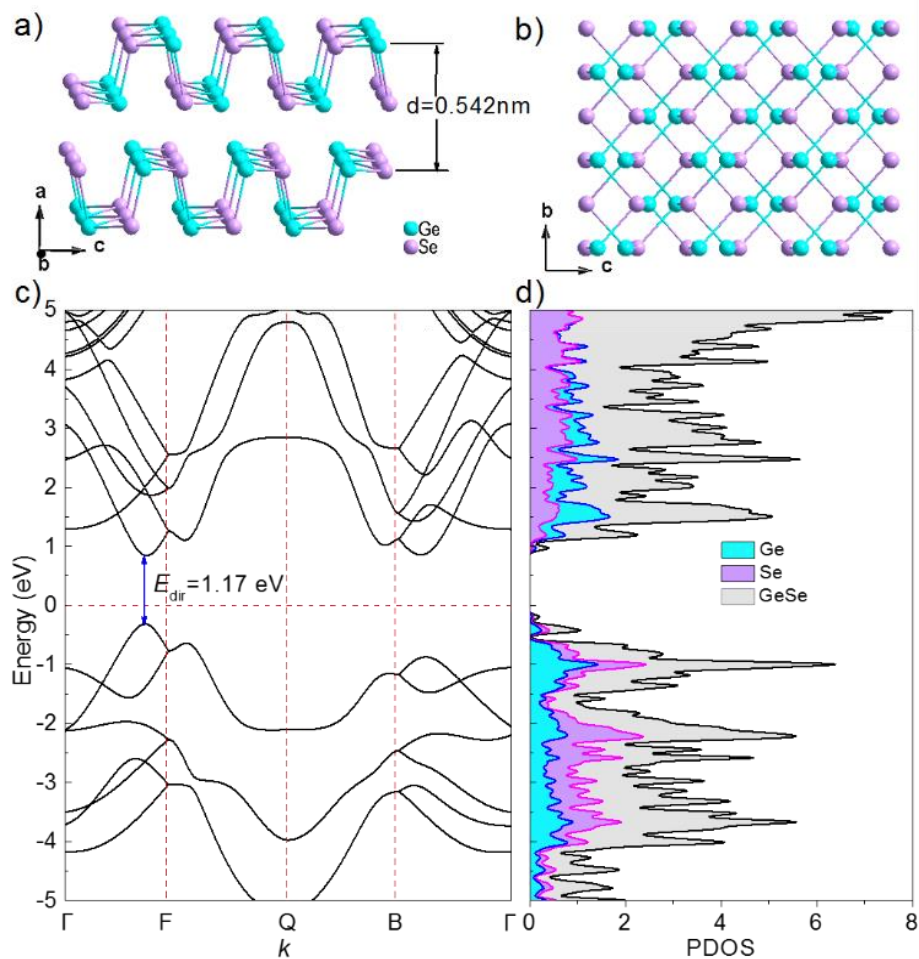

**Figure 1.** (a) Crystal structure of orthorhombic GeSe (space group: Pnma). Each unit cell consists of two GeSe monolayers. (b) Top view of GeSe (along a-axis). (c) Energy band structure. The direct transition is located between Γ-F with an $E_{dir}$ of 1.17 eV. (d) Projected density of states (PODS) for Ge and Se and total DOS for GeSe monolayer.

Stimulated by the peculiar layer-number dependent electronic structure and the presence of a van der Waals gap, synthesis of few (and single) layer GeSe was performed by liquid phase exfoliation (LPE) of bulk GeSe powders (prepared by solid state reaction of elemental powders) in different solvents. The obtained GeSe nanosheets are re-dispersible in different solvents, forming stable colloidal dispersion without noticeable precipitation for over 24 h. Presented in Figure 2 is the GeSe nanosheets exfoliated in the solvent of ethanol under ambient conditions. From the representative transmission electron microscopy (TEM) image, the sheet-like structures with a lateral dimension of 50 - 200 nm are clearly observed. The orthorhombic crystal structure of these FL-GeSe nanosheets is well corroborated by the selected area electron diffraction (SAED) pattern, as well as the high resolution TEM (HRTEM) image (Figure 2c) where lattice spacing of 0.288 nm and 0.279 nm are ascribed to the (011) and (111) planes. The homogeneous elemental distribution of both Ge and Se by a single-sheet elemental mapping together with X-ray photoelectron spectroscopy (XPS) further confirms the chemical characteristic of GeSe nanosheets (Figure 2d-f). Moreover, analysis of these GeSe nanosheets made by atomic force microscopy (AFM) demonstrates an average of thickness of around 2 nm (corresponding to 4 layer stack) with a rather focused thickness distribution over 1.5 – 8 nm (Figure 2g, h). From AFM image, single layer GeSe is present, but obviously is not dominating, which is a general feature of the LPE process.[14]

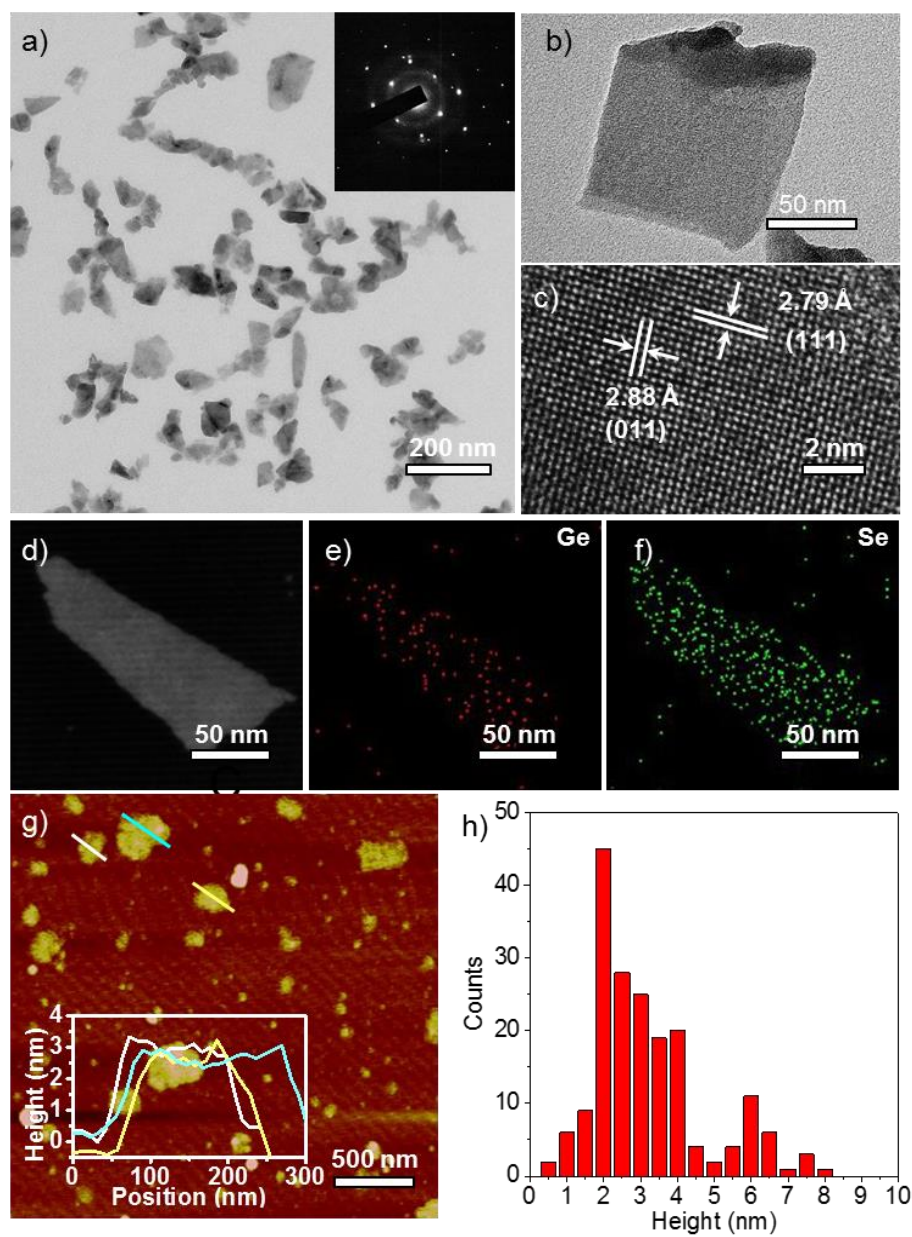

**Figure 2.** (a) Transmission electron micrograph (TEM) and SAED pattern (inset) of the as-synthesized GeSe nanosheets. (b) TEM image of a single GeSe nanosheet. (c) HRTEM images of GeSe nanosheet with clear lattice fringes. (d) Scanning TEM (STEM) image of a single GeSe nanosheet; (e,f) Elemental mapping for the same nanosheet shown in (d). (g)

AFM image and height profile (inset) of the GeSe nanosheets. (h) Statistical analysis (made with over 200 single sheets) for the thickness of the GeSe nanosheets.

To examine the influence of solvent on LPE of GeSe, we have employed a number of solvents with different polarity and surface tension. With appropriate ultrasonic agitation (i.e., time, and intensity) condition, all the solvents examined yield GeSe nanosheets yet with quite different lateral size and thickness. These sheets produced in different solvents are all re-dispersible, forming stable colloidal dispersion with long term stability. The nanosheets exfoliated in 1-Cyclohexyl-2-pyrrolidnone (CHP) show the largest lateral size of up to 400 nm, while the products from isopropyl alcohol (IPA) are dominated by much smaller sheets and nanoparticles. The results therefore suggest that solvent characteristics, such as surface tension, play an important role in LPE of these 2D materials.[15]

Since the stability under ambient condition is of paramount importance for practical application, we then examined the stability of the FL-GeSe thin film (deposited onto quartz substrate) as well as colloidal dispersion, Unlike FL-BP (or phosphorene) that is highly prone to be oxidized,[16] the thin film of FL-GeSe nanosheets stored in ambient condition are stable for over one month without observable degradation as confirmed by Raman spectroscopy. In addition, there are also chemically inert in most of the examined organic solvents; while moderate degradation was observed for the FL-GeSe dispersion stored in solvents of ethanol for over two months from TEM images. Examination by elemental analysis and Raman spectroscopy reveals that the needle and dots in the degradation products are Se and Ge, respectively. This results, however, implies a new synthetic route for these elemental nanostructures.

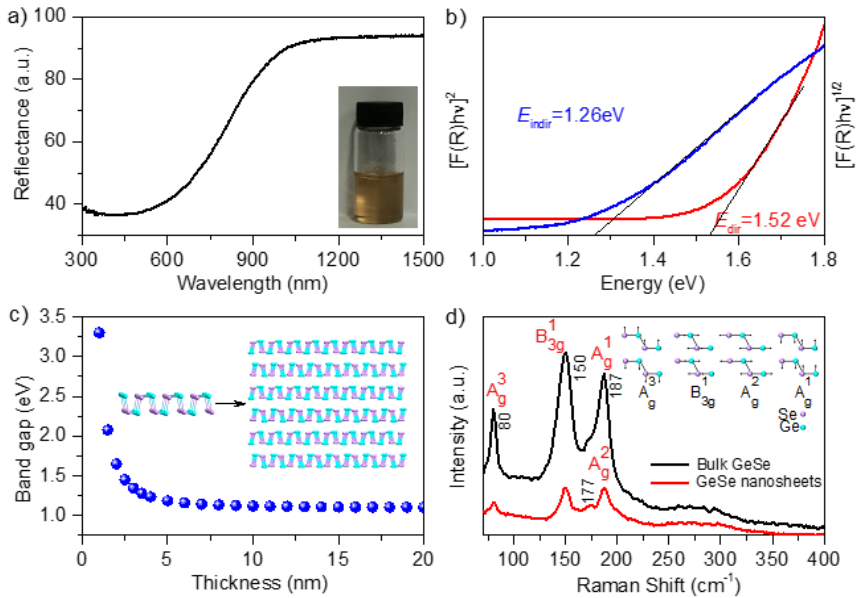

**Figure 3.** (a) Diffuse reflection spectrum of FL-GeSe dispersion spin-coated onto high-purity quartz substrate. Inset shows a photograph of GeSe dispersion in ethanol. (b) Calculation of the direct and indirect bandgap (by using the Tauc plot) from the diffuse reflection spectrum; (c) Dependence of optical bandgap on the thickness of GeSe sheet calculated based on a 2D quantum well model (section 1 in SI). (d) Raman spectra of FL-GeSe nanosheets and bulk GeSe powders. Inset image shows the atomic displacement of the Raman active modes in GeSe.

We determined the optical bandgap of GeSe nanosheets using optical reflectance spectrum, as shown in Figure 3b. After performing Kubelka-Munk transformations, the direct and indirect bandgap found in the Tauc plot are 1.52 eV and 1.26 eV (Figure 3b), respectively, suggesting an indirect bandgap for the FL-GeSe. For bulk GeSe, we found an indirect bandgap of 1.08 eV, slightly lower than that of the FL-GeSe. These results are in agreement with calculation considering that monolayer GeSe in the measured samples are not dominating. In addition, it has to be noted that density functional theory calculation always under-estimate the bandgap and scattering cannot be avoided in optical measurement; these measured bandgaps are therefore reasonable. Furthermore, taking into account that the measured sample consists of FL-GeSe of different layer numbers, the measured bandgap of FL-GeSe is in agreement of the calculated values based on strong quantum confinement effect in 2D systems, where bandgaps of 1.65 and 1.34 eV are predicted for GeSe of thicknesses of 2 nm and 3 nm (Figure 3c).

The thickness-dependent properties were then examined by Raman spectroscopy. For GeSe of the $D_{2h}^{16}$ symmetry, there should be totally 12 active Raman modes ($4A_g+2B_{1g}+4B_{2g}+2B_{3g}$). [17] Here, three vibrational modes (the $A_g^3$ mode at 80 cm$^{-1}$, $B_{3g}^1$ mode at 150 cm$^{-1}$, $A_g^1$ mode at 187 cm$^{-1}$) can be identified in Figure 3d. Compared with bulk GeSe, the peak intensity is notably reduced while the peak intensity remains unchanged. The weakening of Raman peaks is always observed in 2D materials and it can be ascribed to the small sample thickness and presence of rich defects and surface atoms. On the other hand, the reduction of dielectric screening effect in 2D materials normally leads to increase in vibrational frequency, while it is obviously balanced by the weakened short range inter coupling in the FL-GeSe.[18]

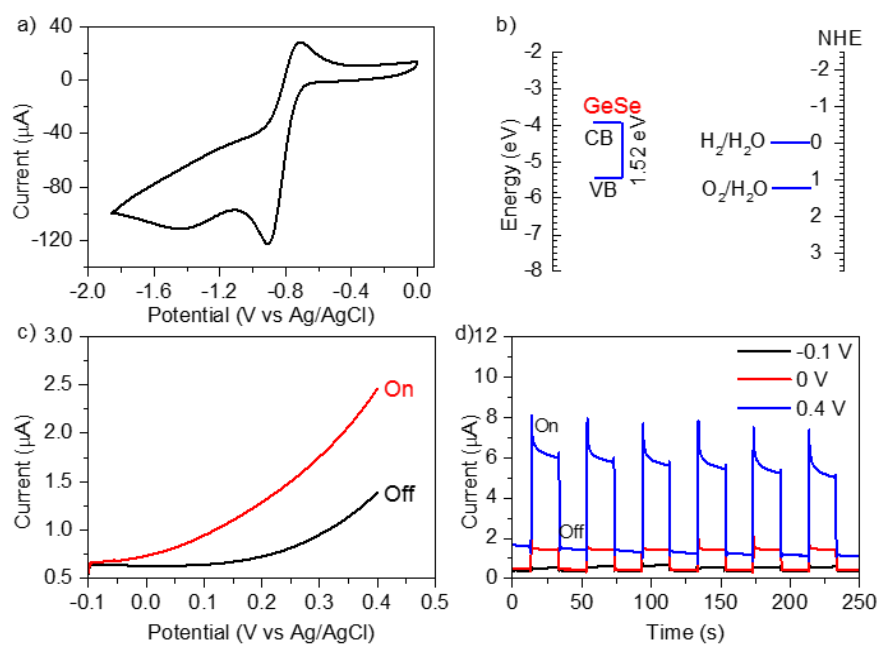

**Figure 4.** (a) Cyclic voltammogram of the GeSe nanosheets recorded in the electrolyte solution of acetonitrile containing 0.1 M tetrabutylammonium perchlorate (TBAP). (b) The position of conduction band minimum (LUMO) and valence band maximum (HOMO) for FL-GeSe. For reference, the positions of hydrogen evolution and oxygen evolution energies are shown. (c) I-V curves of the GeSe nanosheets deposited onto an ITO-coated glass recorded in the presence (on) and absence (off) of excitation from a Hg lamp as the simulated visible light source. (d) Time-dependent photocurrent response of the GeSe nanosheets at bias potential of 0.4, 0, and -0.1 V with light on/off (chopped manually with a time interval of 20 s).

To locate the CB minimum and VB maximum, we recorded cyclic-voltammetry for FL-GeSe deposited onto a glassy carbon electrode (Figure 4a). It is shown that the onset of the reduction potential appears at -0.74 eV. Accordingly, the bottom of the CB (lowest unoccupied molecular orbital, LUMO) was calculated to be 3.96 eV and the top of the valence band (highest occupied molecular orbital, HOMO) values was calculated to be 5.48 eV. The LUMO level is slightly higher than that of bulk GeSe powders and that of SnSe nanocrystals and nanosheets.[19]

Due to the favorable optical bandgap (1.43 eV, 806 nm) that are nearly optimal for photovoltaic applications,[20] here, as a proof-of-concept experiment, we demonstrated photocurrent generation under visible light irradiation by using the FL-GeSe as the working electrode in a standard three-electrode setup. As shown in Figure 4c, the potentiodynamic scans indicate strong photoresponse at potential range of -0.1 – 0.4 V (vs. Ag/AgCl). In addition, the near rectangular current-time relation suggests fast photoresponse (Figure 4d), which can completely exclude the contributions of thermal-induced effects as well as photocorrosion. Compared with the bulk GeSe powder, the photoresponse is more prominent with a higher on/off ratio for the FL-GeSe, benefited from structural two-dimensionality. Probably due to photo-induced charging under constant applied potential, [21] for each pulse an initial high current density is generated upon switching on the light, followed by rapid current stabilization within less that 1s. These results, together with theoretical calculation, suggest that 2D GeSe could be a new photo-active medium for energy-related applications.

In summary, we have shown that one of the phosphorene analogs, few-layer GeSe, can be easily produced by exfoliation and its dispersions and thin film are stable at ambient environment. DFT calculations together with optical characterizations suggest that FL-GeSe shows favorable optoelectronic properties, which are promising for diverse applications. Particularly, our preliminary results have showed that FL-GeSe could be a highly active medium for photocurrent generation under visible light irradiation. The successful experimental access to FL-GeSe enabled by LPE may stimulate further exploration of this phosphorene analogue as well as other group IV mono-chalcogenides for diverse electronic and photonic applications.

**Acknowledgements**

This work is financially supported by the National Natural Science Foundation of China (Grants no. 51132004, 61475047, 11504323). The authors thank Xiaobo Xiang for assitance in the electrochemical measurement and Xiaokun Ding, Qiaohong He, Xinting Cong for TEM observation.